\documentclass{article}

\usepackage{arxiv}

\usepackage[utf8]{inputenc} 
\usepackage[T1]{fontenc}    
\usepackage{hyperref}       
\usepackage{url}            
\usepackage{booktabs}       
\usepackage{amsfonts}       
\usepackage{nicefrac}       
\usepackage{microtype}      
\usepackage{lipsum}
\usepackage{graphicx}
\graphicspath{ {./images/} }
\usepackage{epstopdf, epsfig}
\usepackage{caption}
\usepackage{color, soul}
\usepackage[colorinlistoftodos]{todonotes}

\usepackage{longtable} 
\usepackage{multirow} 
\usepackage{amsmath} 
\usepackage[T1]{fontenc} 
\usepackage{tabularx} 
\usepackage{changepage} 
\usepackage{natbib} 
\usepackage{float} 
\usepackage[running]{lineno} 
\usepackage{xcolor} 
\usepackage{relsize} 
\usepackage{url} 
\usepackage{appendix} 
\title{\rm{On the large-scale vertical velocity intermittency of turbulent wall flows}}

\author{
 Tirtha Banerjee \\
  Department of Civil and Environmental Engineering\\
  Department of Earth System Science\\
  University of California, Irvine\\
  Irvine, CA 92697, USA\\
  *Corresponding author
  (\texttt{tirthab@uci.edu}) \\
   \And
 Elia Buono \\
Dipartimento di Ingegneria dell'Ambiente\\ del Territorio e delle Infrastrutture\\ Politecnico di Torino, Torino, Italia \\
\And
  Costantino Manes\\
  Dipartimento di Ingegneria dell'Ambiente\\ del Territorio e delle Infrastrutture\\ Politecnico di Torino, Torino, Italia\\
\And
Michael Heisel\\
School of Civil Engineering\\ University of Sydney\\ Sydney, Australia\\
\And
Davide Poggi\\
Dipartimento di Ingegneria dell'Ambiente\\ del Territorio e delle Infrastrutture\\ Politecnico di Torino, Torino, Italia\\
\And
Cosimo Peruzzi\\
Area for Hydrology, Hydrodynamics, Hydromorphology and Freshwater Ecology\\Italian Institute for Environmental Protection and Research (ISPRA)\\ Rome, Italy\\
\And
Einara Zahn\\
Department of Earth and Environmental Sciences\\ University of Pennsylvania\\ Philadelphia, PA, USA\\
\And
Elie Bou-Zeid\\
Department of Civil and Environmental Engineering\\ Princeton University\\ Princeton, NJ, USA\\
\And
Gabriel Katul\\
Department of Civil and Environmental Engineering\\ Duke University, Durham, NC, USA\\
The Department of Civil, Construction, and Environmental Engineering\\University of Alabama\\Tuscaloosa, AL, USA\\
}

\begin{document}
\maketitle

\begin{abstract}
Large-scale intermittency in the vertical velocity (LSI) has received significant attention in studies of coherent structures and their detection using data-driven approaches. However, a theory that predicts the origin of LSI from the Navier-Stokes equations or some approximated version of them at very high Reynolds numbers is yet to be achieved.  This letter proposes such a theory for a neutrally stratified wall-bounded turbulent flow based on a dominant balance between inertial and pressure forces. Using multiple flume and wind tunnel experiments, it is shown that the flatness factor ($FF_w$) measuring LSI collapses to a universal trend for all flow configurations within the inertial
sublayer (ISL) before reaching a common minimum value above the ISL.  A theory that predicts $FF_w$ using second-order statistics and explicitly accommodates large-scale energy anisotropy is tested against a wide range of Reynolds numbers from laboratory to field settings with varied surface roughness conditions. The theory also demonstrates why $FF_w$ cannot be described using down-gradient closure approximations routinely employed in large-scale meteorological and climate models. 
\end{abstract}


\section{Introduction}
\label{sec:introduction}
Large-scale intermittency (LSI) in the vertical velocity ($w$) component of wall-bounded high Reynolds number flows is necessary to describe transport across the wall-normal direction by energy-containing eddies. LSI is of significance to vertical momentum and heat exchanges with surfaces, pollutant dispersion, sediment movement and suspension, among others. These processes depend disproportionately on rare but intense wall-normal velocity events \citep{Davidson2015}. The rarity of such events is usually measured by the flatness factor ($FF_w$) of $w$ defined by $FF_w={\overline{w'^4}}/{\sigma_w^4}$, where overline is averaging over coordinates of statistical homogeneity approximated by temporal averaging in many laboratory and field experiments.  Here, velocity fluctuations from their time-averaged values are indicated by primed quantities, and the variance of a turbulent flow variable $s'$ is defined as $\sigma_s^2=\overline{s'^2}$. Extremes in $w'$ bursts are compared to their Gaussian tail expectations using deviations of $FF_w$ from 3.  The $FF_w$ measuring LSI is a property of production-range outer eddies, and differs from the much studied fine-scale intermittency that is a property of the inertial and dissipation range eddies  \citep{sandborn1959measurements,gurvich1967breakdown,frisch1978simple}. At fine or Kolmogorov micro-scales, increases in intermittency are attributed to the role of viscosity in damping less energetic vertical velocity increments and allowing only the most energized events to survive a viscosity censoring mechanism \citep{batchelor1949nature,kuo1972experiment}. On the other hand, energy-containing eddies in $w$ encode key signatures of coherent structures, up-drafts/down-draft events, or bursts in an otherwise less active or quiescent state. There have been recent advances in identifying connections among signatures of coherent structure in time series, LSI, and gradual transitions from LSI to fine-scale intermittency \cite{chowdhuri2021visibility, chowdhuri2023revisiting,chowdhuri2024level}. Such approaches utilize the local topology of the flow to detect LSI and remain diagnostic - not prognostic - by design.  In canonical wall-bounded turbulent flows at very high Reynolds number, it may be conjectured that $FF_w$ is controlled by inertial and pressure forces instead of viscous forces. Yet, a physics-based theory for $FF_w$ that, at minimum, explains its variations in the wall-normal direction ($z$) remains incomplete even for the most idealized flow conditions. It is this gap in LSI that motivates the work here.  
\begin{figure}
    \centering \includegraphics[width=1\linewidth]{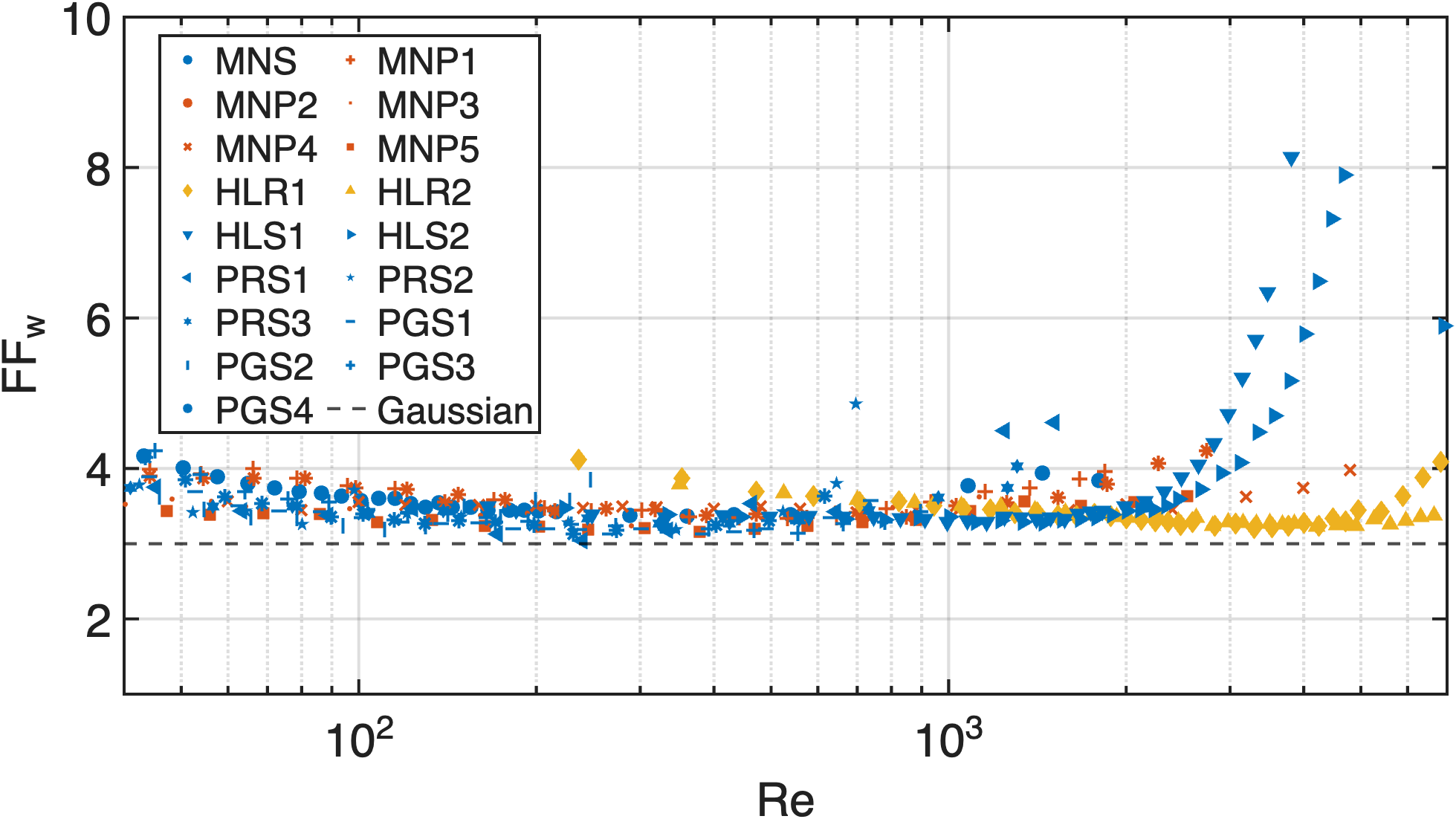}
    \caption{Variation of $FF_w$ against the distance from the wall in inner units ($z^+=u_* z/\nu=Re$) approximating a local Reynolds number $Re$ across a wide corpus of laboratory experiments (Open channel: OC and Wind Tunnel: WT) compiled by \citet{buono2024verticalPoF} and described in Table\ref{tab:lab_summ}. MN  stands for \citet{manes2011turbulent}, HL stands for \citet{heisel2020velocity}, PR stands for \citet{peruzzi2020scaling}, PG stands for \citet{poggi2002experimental}. S, R, and P stands for smooth, rough and porous bed types and the numbers denote different experimental runs.}
    \label{fig:lab_data}
\end{figure}
That such a theory exists is inspired by the unexpected collapse to a quasi-constant $FF_w$ with normalized wall normal distance $z^+=u_*z/\nu$ across multiple laboratory experiments featured in Figure \ref{fig:lab_data}, where $u_*=(\tau_o/\rho)^{1/2}$ is the friction velocity, $\tau_o$ is the wall (or ground) stress, $\nu$ and $\rho$ are the fluid kinematic viscosity and density, respectively. Offering a physics-based explanation for this collapse in Figure \ref{fig:lab_data}, the near independence of $FF_w$ from $z^+$, and why the near-minimum $FF_w$ seems to occur at the transition between the inertial sublayer (ISL) and the outer layer is the main objective of this letter.

\begin{table}[ht]
\centering
\begin{tabular}{l l c c c c c}
\hline
Source & Dataset & Bed & Flow & $\delta \times 10^{-3}$ (m) & $u_* \times 10^{-3}$ ($\mathrm{ms}^{-1})$   & $Re_{\tau}$ \\
\hline
\multirow{6}{*}{\mbox{\citet{manes2011turbulent}}} 
 & MNS  & S & OC & 60  & 41 & 2160 \\
 & MNP1 & P & OC & 96  & 28 & 2349 \\
 & MNP2 & P & OC &110  & 34 & 3234 \\
 & MNP3 & P & OC &115  & 18 & 1856 \\
 & MNP4 & P & OC &146  & 46 & 5840 \\
 & MNP5 & P & OC & 89  & 49 & 3848 \\
\hline
\multirow{4}{*}{\mbox{\citet{heisel2020velocity}}}
 & HLR1 & R & WT &408 &370 & 9611 \\
 & HLR2 & R & WT &391 &550 &13683 \\
 & HLS1 & S & WT &222 &260 & 3681 \\
 & HLS2 & S & WT &203 &350 & 4536 \\
\hline
\multirow{3}{*}{\mbox{\citet{peruzzi2020scaling}}}
 & PRS1 & S & OC &200 &10 &1730 \\
 & PRS2 & S & OC &120 & 8 & 795 \\
 & PRS3 & S & OC & 85 &22 &1657 \\
\hline
\multirow{4}{*}{\mbox{\citet{poggi2002experimental}}}
 & PGS1 & S & OC & 50 &21 &1071 \\
 & PGS2 & S & OC & 45 & 7 & 331 \\
 & PGS3 & S & OC & 42 &30 &1232 \\
 & PGS4 & S & OC & 46 &19 & 845 \\
\hline
\end{tabular}
\caption{Summary of laboratory experiments featured in Figure \ref{fig:lab_data} where $S$ is smooth, $P$ is porous, $R$ is rough,  $\delta$ is the water or boundary layer depth, and $u_*$ is the friction velocity based on wall stress.  OC indicates open channel flow, whereas WT indicates wind tunnel.  The friction Reynolds number $Re_\tau=u_* \delta/\nu$ is also presented, where $\nu$ is the kinematic viscosity. Details about the measurements and setup are featured elsewhere \citep{buono2024vertical,buono2024verticalPoF}.}
\label{tab:lab_summ}
\end{table}

\section{Theory}
\label{sec:theory}
The cartesian coordinate system used here sets $x=x_1$, $y=x_2$, and $z=x_3$ along the longitudinal, lateral, and vertical or wall-normal directions, respectively, with $z=0$ being at the wall, $z=\delta$ being at the top of the turbulent boundary layer, and the instantaneous velocity components along $x$, $y$, and $z$ directions are labeled as $u=u_1$, $v=u_2$, and $w=u_3$, respectively, with $U=\overline{u}$ defining the mean longitudinal velocity. The initial focus is on deriving $FF_w$ versus wall normal distance $z^+$ in stationary and planar homogeneous flow in the absence of subsidence at high Reynolds numbers.  For these idealized conditions, classical phenomenological turbulence models represent higher-order statistics such as $FF_w$ using a down-gradient diffusion closure given as \citep{launder1975progress,deardorff1978closure}
\begin{align*}
FF_w=-\frac{K_t}{\sigma_w^4} \frac{\partial \overline{w'^3}}{\partial z},
\end{align*}
where $K_t=C_1 \sigma_w^2 \tau$ is an eddy diffusivity related to $\sigma_w^2$ and a characteristic time scale $\tau$ (to be discussed later on) using a similarity coefficient $C_1$. While such closure types for high-order moments are in use within community-based geophysical flows \citep{mellor1982development,thayer2015unified}, this model is problematic. Studies have shown $Sk_w$ is constant in the ISL of neutrally stratified boundary layers \citep{buono2024vertical,buono2024verticalPoF}, meaning that gradient diffusion incorrectly predicts zero flatness factor.  

Another common class of models are based on realizability constraints whereby two random variables $a'=w'$ and $b'=w'w'$ must satisfy the Cauchy–Schwarz inequality \citep{Alberghi2002,maurizi2006dependence,buono2024verticalPoF} $\overline{a'b'}\le \sigma_a \sigma_b$. Setting $\sigma_a=\sigma_w$ and $\sigma_b=\sigma_{w^2}=\left[\overline{(b'-\overline{b'})^{2}}\right]^{1/2}=\left[\overline{(w'^2-\overline{w'^2})^2}\right]^{1/2}$ and expanding this expression yields $Sk_w={{\overline{w'w'w'}}}/{\sigma_w^3}\le  \left(FF_w-1\right)^{1/2}$ or  $Sk_w^2\le FF_w-1$, where $Sk_w$ is the vertical velocity skewness. When such inequality constraint is written as an equality with an unknown coefficient, it enables an estimate of $FF_w=\alpha_g (Sk_w^2+1)$ where $\alpha_g$ is a model parameter that should exceed unity to ensure realizability. For a Gaussian PDF, the $Sk_w=0$, $FF_w=3$, and a possible choice for the parameter is $\alpha_g=3$.  Empirical values for $\alpha_g$ ranging from $2.6-3.3$ have been reported across a number of field experiments and Large Eddy Simulations (LES) \citep{buono2024verticalPoF}. The realizability argument with equality replacing inequality can provide empirical justification for a statistical coordination between asymmetry (or $Sk_w$) and LSI (or $FF_w$) in $w'$. These two non-physics-based models of $FF_w$ are revealing: in one, $FF_w$ is proportional to a gradient of the triple moments, whereas in the other $FF_w$ is proportional to the actual squared value of the triple moment, not its vertical gradient. LES studies \citep{stevens2014large} empirically show that the high-order moment $(\overline{w'^{2p}})^{1/p}$ ($p>1$) profile follows expectations from a straightforward extension of Townsend's attached eddy model (AEM)  \citep{townsend1980structure,marusic2019attached}. While such AEM extension to $FF_w$ may explain its near constant value ($\approx 3.3$) in the ISL with respect to $z$ \citep{woodcock2015statistical}, it does not describe the weak decline in $FF_w$ with $z^+$ or, for that matter, an occurrence of a near minimum in Figure \ref{fig:lab_data} at the transition from the ISL to the outer layer. Perhaps less satisfying is that all these aforementioned expressions for $FF_w$ make no explicit contact with the Navier-Stokes equations.   To derive an expression for $FF_w$ from the Navier-Stokes equations, the instantaneous turbulent velocity $u_i'$ is first considered and is given by

\begin{equation}
\frac{\partial u_i'}{\partial t} + \overline{u_j} \frac{\partial u_i'}{\partial x_j} + u_j' \frac{\partial \overline{u_i}}{\partial x_j} + u_j' \frac{\partial u_i'}{\partial x_j} = -\frac{\partial p'}{\partial x_i}+\nu \frac{\partial^2 u_i'}{\partial x_j^2} + \frac{\partial \overline{u_i'u_j'}}{\partial x_j},
\end{equation}
where $p'$ are pressure perturbations normalized by fluid density assumed constant, and $\nu$ is the kinematic viscosity.  For $i=3$ and  $\nu ({\partial^2 u_i'}/{\partial x_j^2})$ much smaller than the inertial terms (i.e., high Reynolds number for large scales) results in
\begin{equation}
\begin{aligned}
\frac{\partial u_3^{\prime}}{\partial t}+
\left(\overline{u_1} \frac{\partial u_3^{\prime}}{\partial x_1}+\overline{u_2} \frac{\partial u_3^{\prime}}{\partial x_2}+\overline{u_3} \frac{\partial u_3^{\prime}}{\partial x_3}\right)+
\left(u_1^{\prime} \frac{\partial \overline{u_3}}{\partial x_1}+u_2^{\prime} \frac{\partial \overline{u_3}}{\partial x_2}+u_3^{\prime} \frac{\partial \overline{u_3}}{\partial x_3}\right)+\\ \left(u_1^{\prime} \frac{\partial u_3^{\prime}}{\partial x_1}+u_2^{\prime} \frac{\partial u_3^{\prime}}{\partial x_2}+u_3^{\prime} \frac{\partial u_3^{\prime}}{\partial x_3}\right)=-\frac{\partial p^{\prime}}{\partial x_3}+\left(\frac{\partial \overline{u_1^{\prime} u_3^{\prime}}}{\partial x_1}+\frac{\partial \overline{u_2^{\prime} u_3^{\prime}}}{\partial x_2}+\frac{\partial \overline{u_3^{\prime} u_3^{\prime}}}{\partial x_3}\right).
\end{aligned}
\end{equation}

Multiplying both sides with ${u_3'}^3$ and averaging results in:
\begin{equation}
\begin{aligned}
\overline{{u_3'}^3\frac{\partial u_3^{\prime}}{\partial t}}+
\left(\overline{{u_3'}^3\overline{u_1} \frac{\partial u_3^{\prime}}{\partial x_1}}+\overline{{u_3'}^3\overline{u_2} \frac{\partial u_3^{\prime}}{\partial x_2}}+\overline{{u_3'}^3\overline{u_3} \frac{\partial u_3^{\prime}}{\partial x_3}}\right)+
\left(\overline{{u_3'}^3 u_1^{\prime} \frac{\partial \overline{u_3}}{\partial x_1}}+\overline{{u_3'}^3 u_2^{\prime} \frac{\partial \overline{u_3}}{\partial x_2}}+\overline{{u_3'}^3 u_3^{\prime} \frac{\partial \overline{u_3}}{\partial x_3}}\right)+\\
\left(\overline{{u_3'}^3 u_1^{\prime} \frac{\partial u_3^{\prime}}{\partial x_1}}+\overline{{u_3'}^3 u_2^{\prime} \frac{\partial u_3^{\prime}}{\partial x_2}}+{\overline{{u_3'}^3 u_3^{\prime} \frac{\partial u_3^{\prime}}{\partial x_3}}}\right)\ =
-\overline{{u_3'}^3 \frac{\partial p^{\prime}}{\partial x_3}}+\left(\overline{{u_3'}^3 \frac{\partial \overline{u_1^{\prime} u_3^{\prime}}}{\partial x_1}}+\overline{{u_3'}^3 \frac{\partial \overline{u_2^{\prime} u_3^{\prime}}}{\partial x_2}}+\overline{{u_3'}^3 \frac{\partial \overline{u_3^{\prime} u_3^{\prime}}}{\partial x_3}}\right).
\end{aligned}
\end{equation}
Assuming a stationary ($\partial (.)/\partial t=0$) and planar homogeneous ($\partial (.)/\partial x_1=\partial (.)/\partial x_2=0$) flow in the absence of subsidence ($\overline{u_3}=0$), multiplying by 5, and then switching to meteorological notation for notational simplicity leads to 
\begin{equation}
5\overline{w'^4 \frac{\partial w'}{\partial z}}=\underbrace{\overline{\frac{\partial {w'}^5}{\partial z}}}_{\rm{Inertia}}= 5 \underbrace{\left (\overline{-{w'}^3 \frac{\partial p'}{\partial z}}\right)}_{\rm{Pressure-Velocity}}+5\underbrace{\overline{{w'}^3} \frac{\partial \overline{w'w'}}{\partial z} }_{\rm{Inertia}}.
\end{equation}
Up to this point, the $FF_w$ is implicit and cannot be determined from this force balance without further assumptions.  In the ISL and when transitioning to the outer layer with increasing $z$, the $\overline{w'^3}(\partial \sigma_w^2/\partial z)$ term is small - compared to the pressure-velocity interaction term. We confirmed that the scaled term $(\overline{w'^3}/u_*^3)(\partial \sigma_w^2/\partial z)(\delta/u_*^2)\approx 0$ for $z/\delta<0.4$ and $z^+>50$ using the same experimental data reported in Figure \ref{fig:lab_data} (plot not shown). Accepting this approximation momentarily leads to the simplified balance
\begin{equation}
-\overline{{w'}^3\frac{\partial p'}{\partial z}}=\frac{1}{5}\overline{\frac{\partial {w'}^5}{\partial z}}.
\label{p_v_eqn}
\end{equation}
Both terms require closure to determine $FF_w$, and conventional schemes are used as logical starting points.

\subsection{An Extended Rotta Closure for the Pressure-Velocity Interaction}
To model the pressure velocity interaction term, the conventional Rotta model is to be extended to higher-order statistics. Upon ignoring the pressure diffusion term (i.e. $\partial \overline{w'p'}/\partial z=0$), the basic Rotta model is given by \citep{Rotta1951,launder1975progress,bou2018role}
\begin{equation}
 -\overline{w'\frac{\partial p'}{\partial z}}=+\overline{p'\frac{\partial w'}{\partial z}} =\frac{C_R}{2\tau}\left(\frac{1}{3}\overline{q}-\overline{w'w'}\right),
\end{equation}
where $q=2K$, $K$ is the instantaneous turbulent kinetic energy ($K=u'u'+v'v'+w'w'$), $C_R=1.8$ is the Rotta constant, and $\tau$ is a height-dependent relaxation time scale formed from $K$ and the mean turbulent kinetic energy dissipation rate. This conventional closure scheme applies to many types of turbulent flows. It suggests that the pressure-velocity interaction term is expected to be positive when $\sigma_w^2<(1/3) \overline{q}$, which is confirmed by experiments and simulations \citep{bou2018role}.  Moreover, wall-blocking effects are expected to act in the opposite direction, thereby ameliorating the effectiveness of the pressure distribution near boundaries \citep{launder1975progress,mccoll2016mean}. 

To extend the simplified Rotta scheme to the application at hand, the approach used for third order statistics is followed \citep{buono2024vertical}.  This approach leads to a straight forward extension requiring the multiplication of $w'^2$ on both sides of the Rotta scheme prior to averaging and results in 
\begin{equation}
    -\overline{{w'}^3\frac{\partial p'}{\partial z}}=\frac{C_R}{2\tau}\left(\frac{1}{3}\overline{{w'}^2{q}}-\overline{{w'}^4}\right).
\end{equation}
In this derivation, interactions between $\tau$, $K$ and other velocity terms are ignored.  Thus, the overall balance reduces to
\begin{equation}
\frac{2 \tau}{5 C_R}\frac{\partial \overline{w'^5}}{\partial z}=\frac{2}{3}\overline{w'^2(u'u'+v'v'+w'w')}-\overline{w'^4}=\frac{2}{3}\overline{w'^2(u'u'+v'v')}-\frac{1}{3}\overline{w'^4}.
\end{equation}

\subsection{Quasi-Gaussian Approximations (QGA)}
In the Quasi-Gaussian Approximation (QGA), where some deviations from a strict Gaussian PDF for $w'$ are allowed (e.g. $Sk_w\neq0$ and $FF_w\neq3$), it is assumed that for any four random variables ($a'$, $b'$, $c'$, and $d'$) \citep{millionshchikov1941theory}
\begin{align*}
\overline{a'b'c'd'}=\overline{a'b'}~ \overline{c'd'}+\overline{a'c'}~ \overline{b'd'}+\overline{a'd'}~ \overline{b'c'}.
\end{align*}
With this approximation,
\begin{align*}
\overline{w'^2(u'u'+v'v')}=2\left( \overline{w'u'} ~\overline{w'u'}+ \overline{w'v'} ~\overline{w'v'}\right)+\left(\overline{w'w'}~\overline{u'u'}+\overline{w'w'}~\overline{v'v'}\right)=2u_*^4+\sigma_w^2 \left(\sigma_u^2+\sigma_v^2\right).
\end{align*}
Applying the QGA for the overall balance leads to
\begin{equation}
\frac{1}{3}\overline{w'^4}=\underbrace{\frac{2}{3}\left[2u_*^4+\sigma_w^2 \left(\sigma_u^2+\sigma_v^2\right)\right]}_{\rm{Pressure~Velocity}}-\underbrace{\frac{2 \tau}{5 C_R}\frac{\partial \overline{w'^5}}{\partial z}}_{\rm{Inertia}}.
\end{equation}
 
\subsection{A Model for the Inertial Term}
An estimate of the inertial term may be conducted by assuming that 
\begin{equation}
\frac{1}{5} \frac{\partial \overline{w'^5}}{\partial z}=\overline{\left(w'^4\right) \left(\frac{\partial w'}{\partial z}\right)}=B_L ~ \overline{w'^4} ~ \left(\frac{\sigma_w}{l_o}\right),
\end{equation}
where $l_o$ is a characteristic length scale.  Here, it may be argued that the magnitude of $w'$ scales with $\sigma_w$. There are two end-member choices for $l_o$: (i) a macro-scale choice $L_I$ where $l_o=L_I=\tau \sigma_w$ and this choice may be deemed consistent with the time scale of the Rotta scheme and (ii) a small-scale choice such as the Taylor micro-scale $l_o=\lambda$. The Taylor micro-scale and $L_I$ are related using $\lambda/L_I\propto R_L^{-1/2}$, where $R_L=\sigma_w L_I/\nu$ is another large-scale Reynolds number \citep{tennekes1972first}. For $l_o=\lambda$, it is clear that $B_L$ is no longer a constant but varies with $R_L$.  If $\lambda$ is replaced by an even finer length, the Kolmogorov microscale $\eta$ where viscous effects are significant, then $\eta/L_I \sim R_L^{-3/4}$ \citep{tennekes1972first} also implying that $B_L$ is not a constant but varies with a Reynolds number.  For the macro-scale choice $l_o=L_I=\tau \sigma_w$, the inertial term is
\begin{equation}
\frac{1}{3}\left(1+\frac{6B_L}{C_R}\right)\overline{w'^4}=\frac{2}{3}\left[2u_*^4+\sigma_w^2 \left(\sigma_u^2+\sigma_v^2\right)\right].
\end{equation}
Defining the dimensionless velocity scales $A_u^2=\sigma_u^2/u_*^2$, $A_v^2=\sigma_v^2/u_*^2$, and $A_w^2=\sigma_w^2/u_*^2$ yields
\begin{equation}
FF_w=\frac{4}{1+\alpha_I} \left[\frac{1}{A_w^4}+\frac{\left(A_u^2 + A_v^2 \right)}{{2A_w^2}}\right]; \alpha_I=\frac{6B_L}{C_R},
\label{FFw_AEH_eq}
\end{equation}
where $\alpha_I$ is a coefficient of order unity that reflects the role of inertia in ameliorating $FF_w$ in equation \ref{FFw_AEH_eq}.  Its value may be estimated as $\alpha_I=3$ when using typical near-neutral atmospheric surface layer (ASL) values of $FF_w=3.4$, $A_u=2.4$, $A_v=2.1$, and $A_w=1.3$ as surrogates for very high Reynolds number flows \citep{katul1996investigation}. 
\begin{figure}
    \centering
    \includegraphics[width=1\linewidth]{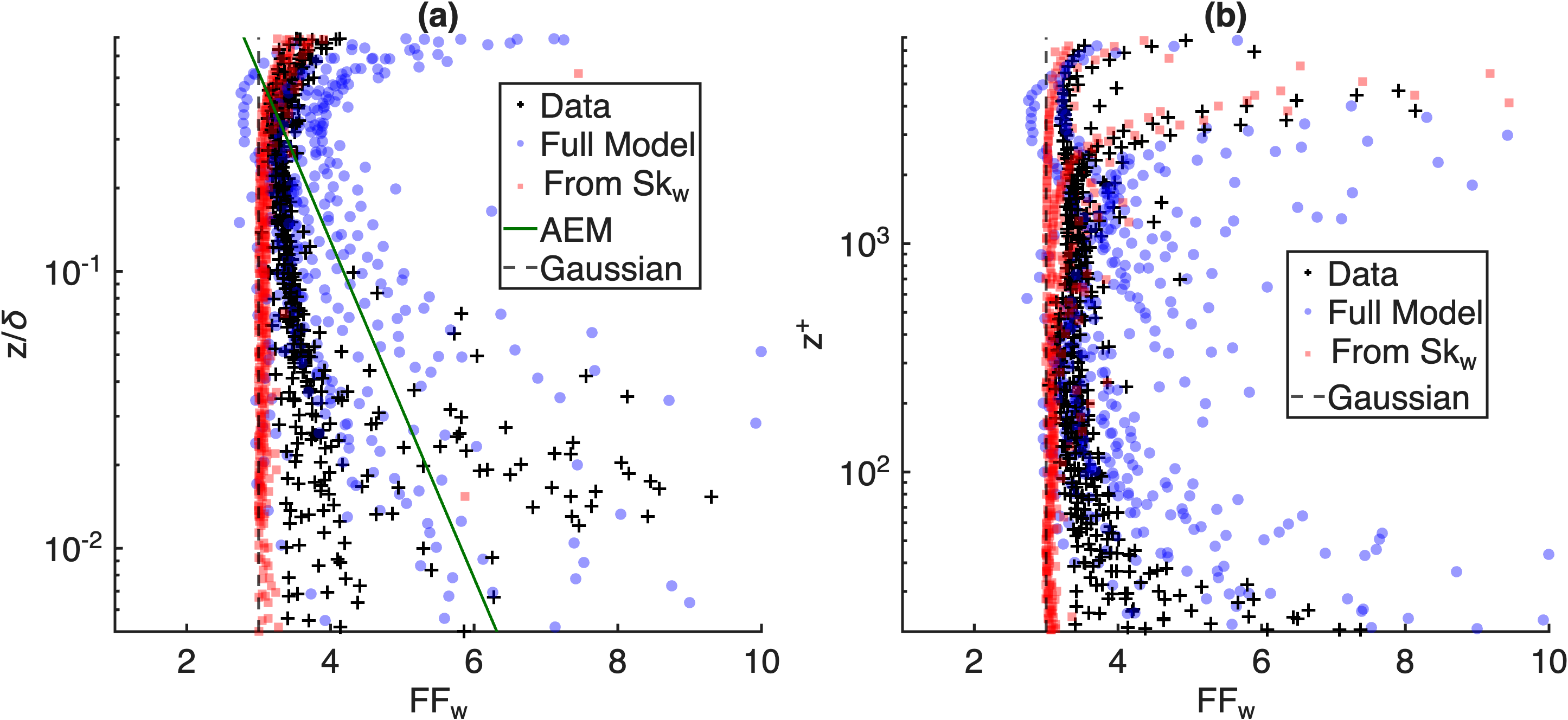}
    \caption{Variation of $FF_w$ with $z/\delta$ (panel a) and $z^+$ (panel b). The $+$ symbols denote the Laboratory observations, the blue circles denote the full model with measured $A_w$ and $A_u$, the red squares denote the Skewness-based prediction, and the green line represents the prediction using the coefficients as fitted for the attached eddy model (AEM). The dashed black line denotes the Gaussian prediction of $FF_w=3$ for both panels.}
    \label{fig:model_comparison}
\end{figure}
Beyond the Rotta and the QGA closure, the derivation adopts two other assumptions:  $\overline{w'^3}(\partial \sigma_w^2/\partial z)$ is small (which is validated as discussed earlier), and $l_o=\tau \sigma_w$. Those assumptions are likely to hold in the ISL and perhaps in transition zones from the buffer layer into the ISL or from the ISL into the outer layer.  However, those two assumptions are unlikely to hold inside the buffer region or well into the outer region.  In keeping with common layer delineations for wall-bounded flows, the buffer region is for $15<z^+<50$, the ISL is for $z^+>100$ and $z/\delta<0.1-0.2$, and the outer layer is defined by $z/\delta>0.5$ \citep{Pope2000}.   
\section{Discussion and Conclusion}
\label{sec:conclusion}
A comparison between measured and modeled $FF_w$ is presented in Figure \ref{fig:model_comparison}. As reference, $FF_w=3$ and $FF_w=\alpha_g (1+Sk_w^2)$ are shown.  Equations $FF_w=\alpha_g (1+Sk_w^2)$ and \ref{FFw_AEH_eq} use the measured $Sk_w$, $A_u$, $A_w$ with $A_v^2=(A_u^2+A_w^2)/2$ while setting $\alpha_I=3$ and $\alpha_g=3$ as constants independent of $z^+$.  The normalized wall normal distances are presented in both outer layer (left) and inner layer (right) variables to highlight trends and model performance in different sub-layers. 
Outside the ISL, deviations from $FF_w=3$ are large hinting that QGA is not valid. The model $FF_w=\alpha_g (1+Sk_w^2)$ captures increases in $FF_w$ with increases in $z/\delta$ reasonably in the outer layer mainly due to the rapid increase in measured $Sk_w^2$ \citep{buono2024verticalPoF}.  However, decreases in $FF_w$ with increases in $z^+$ are not captured by $FF_w=\alpha_g (1+Sk_w^2)$ inside the buffer region. This model also underestimates $FF_w$ in the ISL when setting $\alpha_g=3$. 

When setting $\alpha_I=3$, calculations using equation \ref{FFw_AEH_eq} based on measured $A_u$ and $A_w$ capture the overall profile trends in the $FF_w$ data across much of the buffer and ISL, as well as transitions from the ISL to the outer layer.  To be clear, equation \ref{FFw_AEH_eq} misses contributions from $\overline{{w'}^3} {\partial \overline{w'w'}}/{\partial z}$, which can be large in the buffer- and outer- layers but not in the ISL.  The ${\partial \overline{w'w'}}/{\partial z}$ is positive in the buffer layer but negative in the outer layer, whereas $Sk_w<0$ in the buffer layer and $Sk_w>0$ in the outer layer.  Thus, $\overline{{w'}^3} {\partial \overline{w'w'}}/{\partial z}<0$ and large in magnitude in the buffer and outer layers but near zero in the ISL.  Thus, maximum distortions from $\overline{{w'}^3} {\partial \overline{w'w'}}/{\partial z}$ to model calculations from equation \ref{FFw_AEH_eq} are expected to be far from the ISL and transitions zones to/from ISL. 

\begin{figure}
    \centering
    \includegraphics[width=1\linewidth]{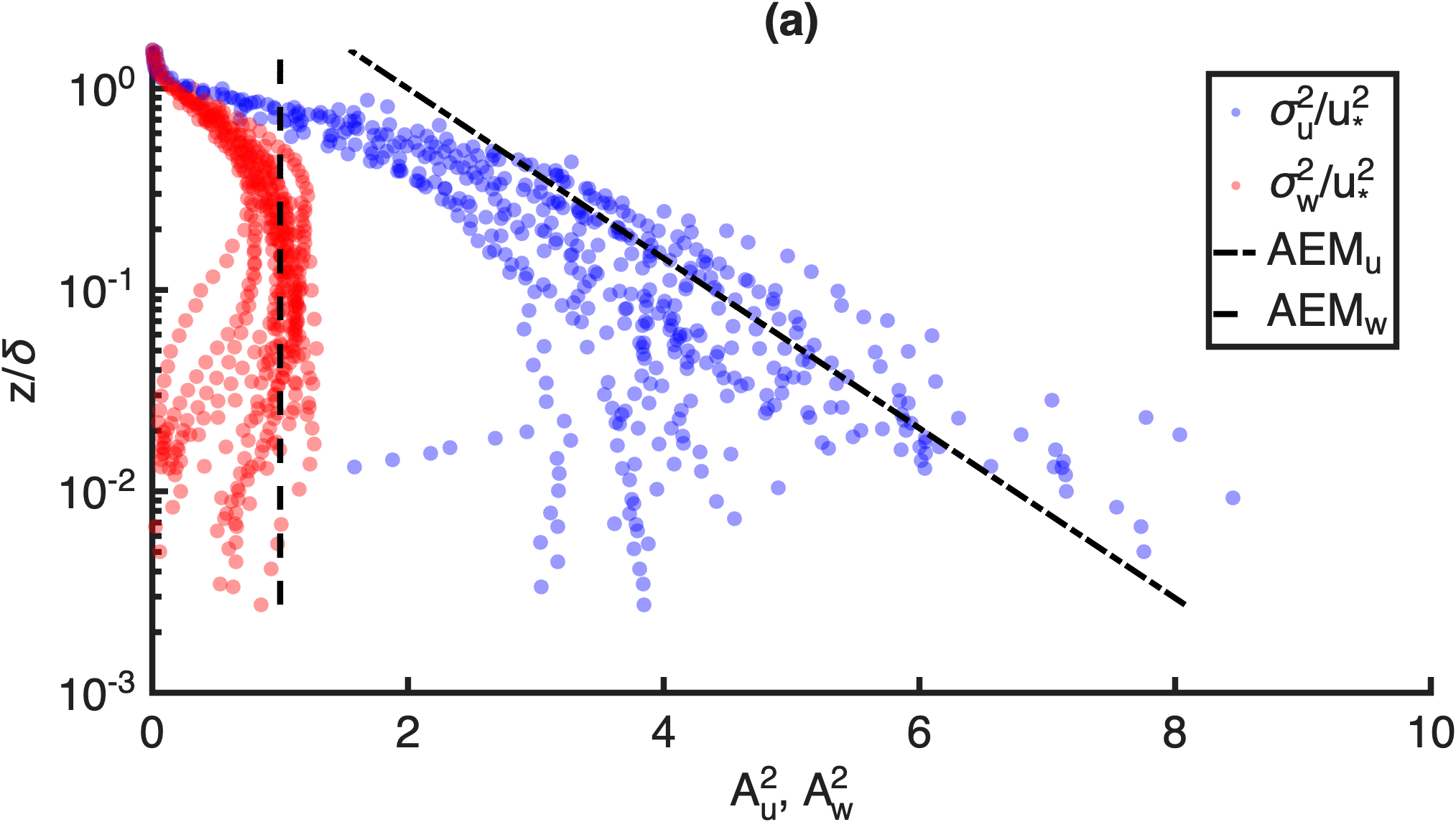}
    \caption{Variation of $A_u$ (blue) and $A_w$ (red) with $z/\delta$ across experiments in Table\ref{tab:lab_summ}, where $\delta$ is the boundary layer or water depth. The AEM predictions using asymptotic values (i.e. large Reynolds numbers) for $A_1=1$, $B_1=2$, and $B_2=1$ are shown (dashed lines) for reference.  These asymptotic values are used in model calculations labeled as AEM.}
    \label{fig:AEM_prediction}
\end{figure}

\begin{figure}
    \centering
    \includegraphics[width=1\linewidth]{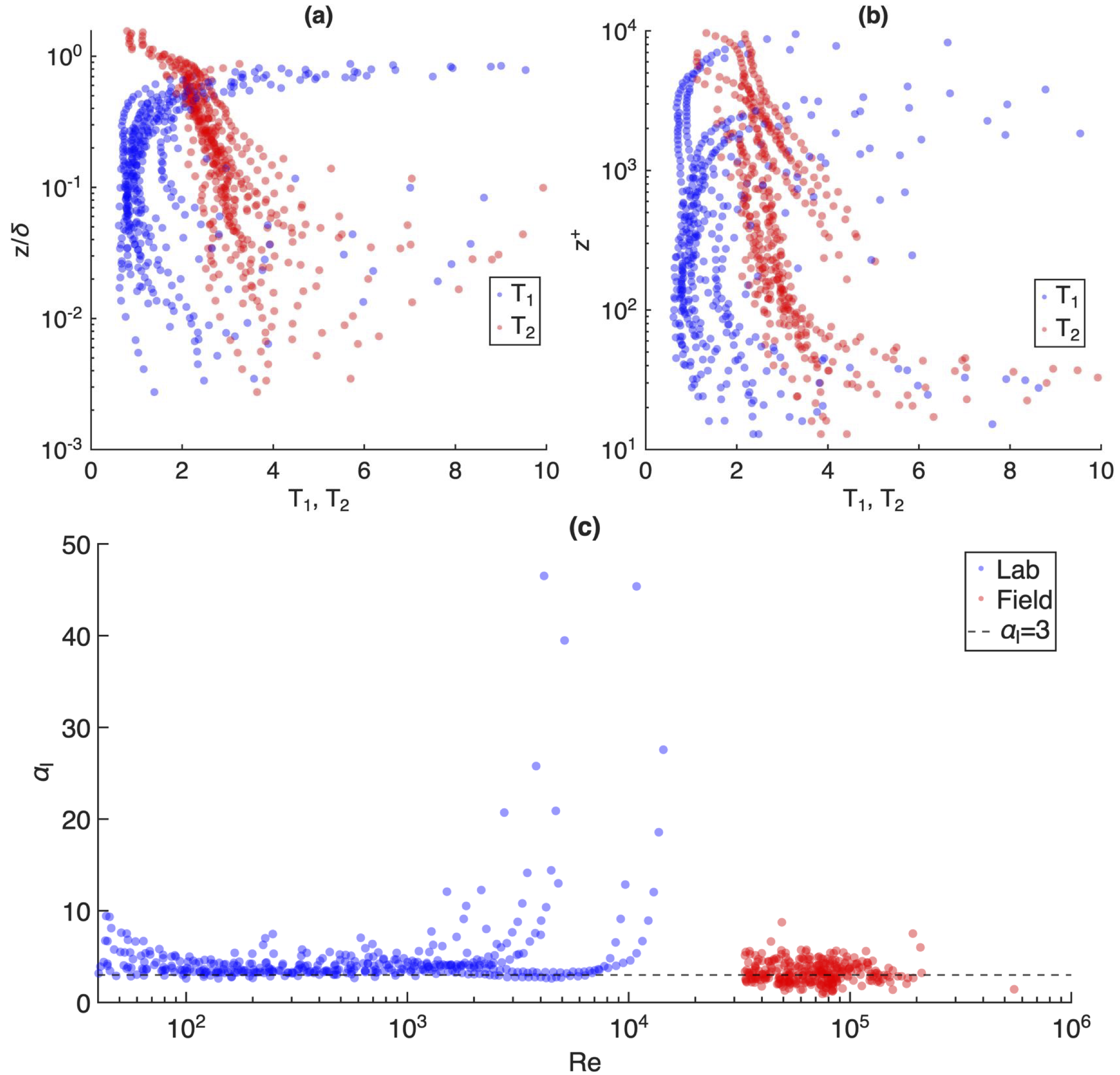}
    \caption{Variation of $T_1=1/A_w^4$ and $T_2=(A_u^2+A_v^2)/(2A_w^2)$ with $z/\delta$ (panel a) and $z^+$ (panel b) across the Laboratory experiments. Panel (c) shows the Estimation of $\alpha_I$ from measured $FF_w$ and $Re=z u_*/\nu=z^+$ for all laboratory (blue) and field experiments (red) taken in the atmospheric surface layer under near-neutral conditions.  The assumed value $\alpha_I=3$ in model calculations is also shown (dashed).}
    \label{fig:PV_terms}
\end{figure}

Equation \ref{FFw_AEH_eq} does not predict $FF_w$ variations with $z$ but a relation between $FF_w$ and normalized second-order flow statistics. To explain patterns of $FF_w$ variations with $z$ requires links between $A_u$ and $A_w$ and $z$. Such a link can be supplied by the AEM but only in the ISL. The AEM for second-order moments predicts 
\citep{townsend1980structure,smits2011high,marusic2019attached}
\begin{align*}
\label{eqn:law_of_wall_2}
 A_u^2= B_1-A_1 \ln\left(\frac{z}{\delta}\right); ~ A_w^2= B_2.
\end{align*}
The $A_1$ is the Townsend-Perry coefficient, and $B_1$ and $B_2$ are coefficients that depend on the flow (e.g., pipe flow versus wind tunnels).  The $A_1$, $B_1$, and $B_2$ are presumed to attain asymptotically constant values at very large Reynolds numbers \cite{townsend1980structure,banerjee2013logarithmic,marusic2013logarithmic,qin2025asymptotic} though in many experiments, these constants differ from the asymptotic limit in numerical values (especially $B_1$ and $A_1$) as shown in Fig. \ref{fig:AEM_prediction} for the experiments in Table\ref{tab:lab_summ}. 
The $FF_w$ calculated using the asymptotic values of the AEM coefficients are also shown in Figure \ref{fig:AEM_prediction} (dashed lines) with $\alpha_I=3$.   The goal of this comparison is not to assess model fidelity but to suggest that the mild decreasing trend in measured $FF_w$ empirically detected in Figure \ref{fig:AEM_prediction} with $z$ can be attributed to an $A_u^2$ dependence on $z/\delta$ predicted by the AEM.  Moreover, equation \ref{FFw_AEH_eq} offers an explanation as to why $FF_w$ is a minimum when transitioning from the ISL to the outer layer. As derived, $FF_w$ is linked to two mechanisms: a $T_1=(1/A_w)^4$ and a large scale energy anisotropy $T_2=(A_u^2+A_v^2)/(2A_w^2)$, both are plotted against $z/\delta$ and $z^+$ in Figure \ref{fig:PV_terms}(a,b).  As expected, $T_2$ largely declines with increasing $z/\delta$ as the flow tends to become energetically more isotropic away from the generation source. The $T_1$ exhibits a minimum in the ISL and increases rapidly as $z/\delta>0.5$.  Because $T_2>T_1$ in the buffer and ISL but $T_1>T_2$ in the upper parts of the outer layer. This explains why $FF_w$ exhibits a local minimum in the transition from the ISL to the outer layer and a weak z-dependence due to the AEM influence on $T_2$ in the ISL.

The remaining `thorny' issue is setting $\alpha_I$ to a constant.  This restrictive assumption is now explored using an independent but expansive data set from the near-neutral ASL.  The data covers $FF_w$ measurements from lakes, bare soil, a grass surface, and two forested sites where $A_u$, $A_v$, $A_w$, and $FF_w$ are measured by triaxial sonic anemometry  \citep{banerjee2024single}.  Using those measurements for near-neutral conditions, the computed $\alpha_I$ from equation \ref{FFw_AEH_eq} is compared against a constant $\alpha_I=3$ in Figure \ref{fig:PV_terms}c. As expected, the ASL data experience some two orders of magnitude higher $z^+$ compared to the laboratory experiments despite that all measurements were conducted in the first few meters above the ground or zero-plane displacement, where $\delta$ may be on the order of a 1000 m.  These findings indirectly confirm that $\alpha_I$ and $B_L$ may be operationally treated as constants when setting $C_R=1.8$.  A constant $\alpha_I$ also implies that $l_o$ is not related to $\lambda$ (or $R_L$) and the choice of $l_o$ as a macro-scale is validated. 

Coordination between $Sk_w$ and $FF_w$ also becomes evident when noting that a similar dominant balance between inertia and pressure redistribution leads to $Sk_w={(2/3)}{(\kappa A_1)}/{A_w^3}$ \citep{buono2024vertical} when $B_2$ is assumed to be independent of $z$ and $\kappa$ is the von Karman constant.  This expression for $Sk_w$ was derived using similar closure models and tested for the same data sets reported here \cite{buono2024vertical,buono2024verticalPoF}.   To conclude, the dominant balance between inertial and pressure–velocity interactions governs large-scale intermittency in $w'$ when quantified by $FF_w$. A theory that represent these two terms captures the observed weak dependence of $FF_w$ on wall-normal distance and flow configuration, highlighting the dominant role of large-scale energy anisotropy and the scaling of turbulent velocity variances. The theory also explains why $FF_w$ has a local minimum at the transition from the inertial to the outer layer.  The introduction of a single similarity parameter $\alpha_I$ provides a practical pathway for modeling higher-order turbulence statistics from lower-order ones beyond traditional closure approaches. These findings clarify the physical origin of large-scale intermittency, explain why conventional gradient-diffusion models fail to reproduce $FF_w$, and offer improved formulations for applications where intermittent vertical transport plays a role. 



\section*{acknowledgments}
TB acknowledges the funding support from the University of California Office of the President (UCOP) grant LFR-20-653572 (UC Lab-Fees); the National Science Foundation (NSF) grants NSF-AGS-2146520, NSF-OISE-2114740, NSF-CPS-2209695, NSF-ECO-CBET-2318718, NSF-RISE-2536815 and NSF‐DMS‐2335847; the United States Department of Agriculture (USDA) grant 2021-67022-35908 (NIFA); and a cost reimbursable agreement with the USDA Forest Service 20-CR-11242306-072. EB acknowledges Politecnico di Torino (Italy) for supporting the visit to Duke University. GK acknowledges support from Los Alamos National Laboratory (USA) through the Strategic Environmental Research and Development Program (SERDP) grant (RC25-0189). DP acknowledges support from Fondo europeo di sviluppo regionale (FESR) for project Bacini Ecologicamente sostenibili e sicuri, concepiti per l'adattamento ai Cambiamenti ClimAtici (BECCA) in the context of Alpi Latine COoperazione TRAnsfrontaliera (ALCOTRA) and project Nord Ovest Digitale e Sostenibile - Digital innovation toward sustainable mountain (Nodes - 4).

\bibliographystyle{apalike}  
\bibliography{turbbib}

\end{document}